\documentclass[
  ]
  {aipproc}

\layoutstyle{6x9}

\begin{document}

\title{The Effect of Positivity Constraints on Polarized\\[2mm]
 Parton Densities}



\author{E. Leader}{
  address={Imperial College London, London WC1E 7HX, England}
}

\author{A.V. Sidorov,}{
  address={Bogoliubov Theoretical Laboratory,
Joint Institute for Nuclear Research, 141980 Dubna, Russia } }

\author{\underline{D.B. Stamenov}}{
%  address={<common address for author2 and author3>}
  address={Institute for Nuclear Research and Nuclear Energy,
1784 Sofia, Bulgaria}
% additional visiting address
}

\begin{abstract}
The impact of positivity constraints on the polarized parton
densities has been studied. Special attention has been paid to
the role of positivity constraints in determining the polarized
strange quark and gluon densities, which are not well determined
from the present data on inclusive polarized DIS.
\end{abstract}

\maketitle

%%%%%%%%%%%%%%%%%%%%%%%%%%%%%%%%%%%%%%%%%%%%
%% MAINMATTER
%%%%%%%%%%%%%%%%%%%%%%%%%%%%%%%%%%%%%%%%%%%%

%\section{Introduction}

Spurred on by the famous European Muon Collaboration (EMC)
experiment \cite{EMC} at CERN in 1987, there has been a huge
growth of interest in the partonic spin structure of the nucleon,
{\it i.e.}, how the nucleon spin is built up out from the
intrinsic spin and orbital angular momentum of its constituents,
quarks and gluons. Our present knowledge about the spin structure
of the nucleon comes from polarized inclusive and semi-inclusive
DIS experiments at SLAC, CERN, DESY and JLab, polarized
proton-proton collisions at RHIC and polarized photoproduction
experiments. The determination of the longitudinal polarized
parton densities in QCD is one of the important aspects of this
knowledge. Many analyses \cite{Groups,LSS05} of the world data on
inclusive polarized DIS have been performed in order to extract
them. It was shown that if the convention of a flavor symmetric
sea is used\footnote{In the absence of polarized charged current
neutrino experiments a flavor decomposition is not possible.} the
polarized valence quarks are well determined, while the polarized
strange sea and polarized gluon densities are weakly
constrained\footnote{About the situation in semi-inclusive DIS
see the talk by R. Sassot at this Workshop \cite {SIDIS}.}.

In this talk we will discuss the effect of positivity constraints
on the polarized parton densities and will demonstrate their
importance in determining the strange and gluon densities,
especially at high $x$.

The polarized parton densities have to satisfy the positivity
condition, which in LO QCD implies:
\begin{equation}
\vert {\Delta f_i(x,Q^2)}\vert \leq f_i(x,Q^2),~~~~ \vert
{\Delta\bar{f_i}(x,Q^2)}\vert \leq \bar{f}_i(x,Q^2). \label{pos}
\end{equation}

The constraints (\ref{pos}) are the consequence of a probabilistic
interpretation of the parton densities in the naive parton model,
which is still valid in LO QCD. Beyond LO the parton densities are
not physical quantities and the positivity constraints on the
polarized parton densities are more complicated. They follow from
the positivity condition for the polarized lepton-hadron
cross-sections $\Delta \sigma_i$ in terms of the unpolarized ones
($\vert {\Delta \sigma_i}\vert \leq \sigma_i$) and include also
the Wilson coefficient functions. It was shown \cite{AFR},
however, that for all practical purposes it is enough, at the
present stage, to consider LO positivity bounds for LO as well as
for for NLO parton densities, since NLO corrections are only
relevant at the level of accuracy of a few percent. Note that, if
the positivity constraints (\ref{pos}) are imposed at some
$Q^2_0$, they are satisfied at any $Q^2 > Q^2_0$ \cite{OT}. So, in
order to control easily the positivity conditions (\ref{pos}) it
is enough to impose them for the minimum value of $Q^2=Q^2_0$ in
the data set used in the QCD analysis.
\begin{figure}
\resizebox{.40\columnwidth}{!}
{\includegraphics{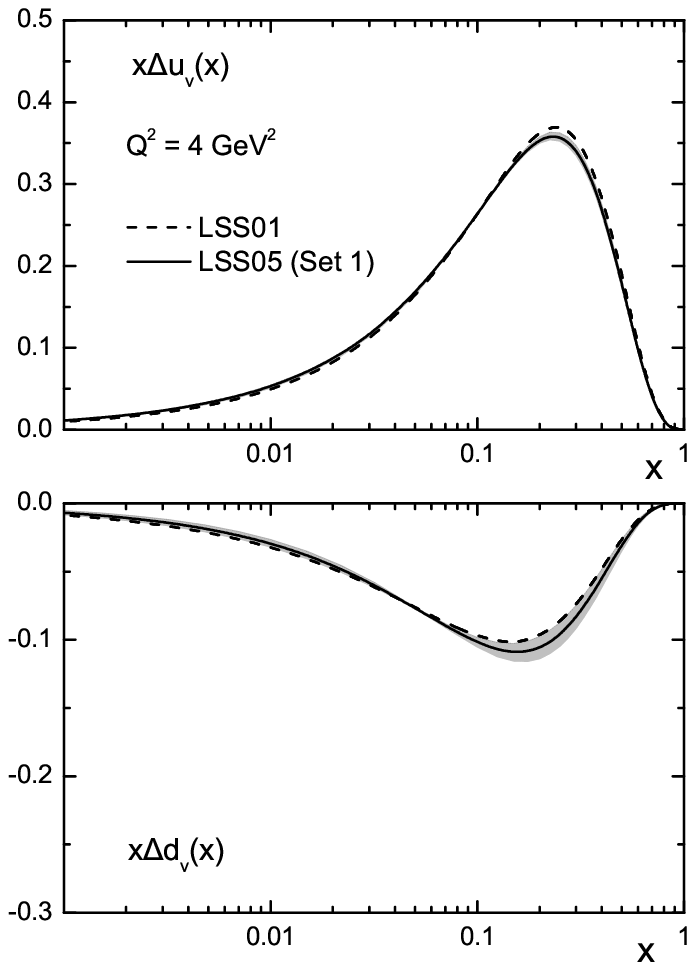}}
\resizebox{.40\columnwidth}{!}
{\includegraphics{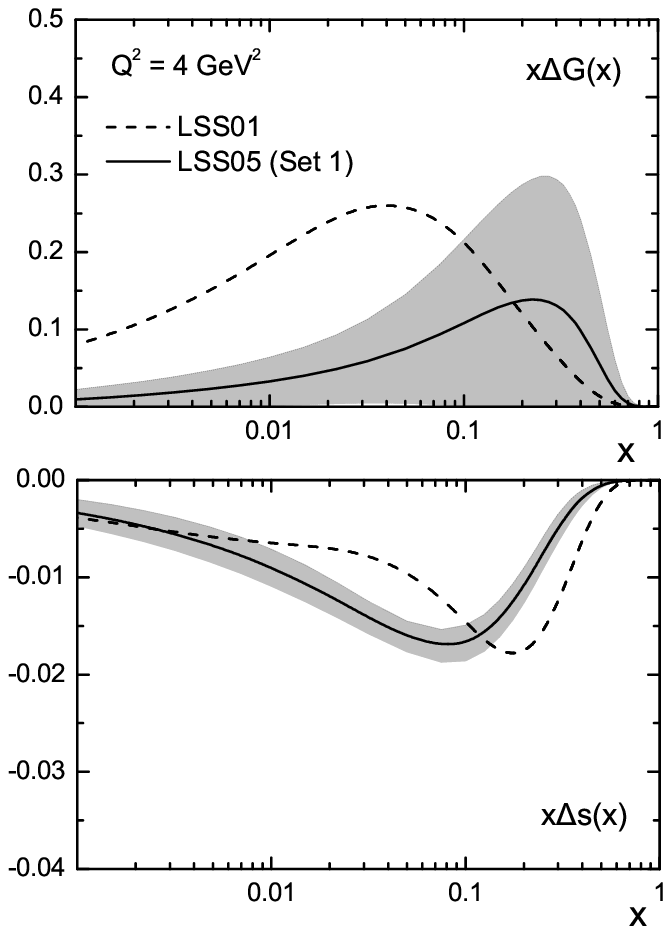}}
  \caption{Comparison between our two sets of NLO(${\rm \overline {MS}}$)
polarized parton densities, LSS'01 and LSS'05(Set 1), at
$Q^2=4~GeV^2$.}
\end{figure}

Let us consider how the use of different positivity constraints
influences the results on the polarized parton densities. In Fig.
1 we compare LSS'05(Set 1) NLO($\rm \overline {MS}$) polarized
parton densities \cite{LSS05} with LSS'01 parton densities
\cite{LSS2001} presented on the HEPDATA web site. Both sets are
determined from the data by the same method but using different
positivity constraints.  While the LSS'05 polarized PD are
compatible with the positivity bounds (\ref{pos}) imposed by the
MRST'02 unpolarized parton densities \cite{MRST02}, those of the
LSS'01 set are limited by the Barone et al. unpolarized parton
densities \cite{Barone}. As seen from Fig. 1, the valence quark
densities $\Delta u_v$ and $\Delta d_v$ of the two sets are close
to each other, while the polarized strange sea quark and gluon
densities are significantly different. This comparison is a good
illustration of the fact that the present inclusive polarized DIS
data allow a much better determination of the valence quark
densities (if SU(3) symmetry of the flavour decomposition of the
sea is assumed) than the polarized strange quarks $\Delta s(x,
Q^2)$ and the polarized gluons $\Delta G(x, Q^2)$. This is
especially true for the high $x$ region, where the values of
$\Delta s(x, Q^2)$ and $\Delta G(x, Q^2)$ are very small and the
precision of the data is not enough to extract them correctly.
That is why different unpolarized sea quark and gluon densities
(see Fig. 2) used on the RHS of the positivity constraints
(\ref{pos}) are important and crucial in determining $\Delta s(x,
Q^2)$ and $\Delta G(x, Q^2)$ in this region. The more restrictive
$s(x, Q^2)_{\rm MRST'02}$ at high $x$ leads to a smaller value of
$\vert {\Delta s(x, Q^2)}\vert_{\rm LSS'05}$ in this region,
while the smaller $G(x, Q^2)_{\rm Bar.et.al}$ provides a stronger
constraint on $\Delta G(x, Q^2)_{\rm LSS'01}$ (see Fig. 1).

To illustrate this fact once more, we compare the LSS'05 (Set 1)
PPD at $Q^2=4~GeV^2$ with those \cite{Groups} obtained by GRSV,
Blumlein, Bottcher and the Asymmetry Analysis Collaboration (AAC)
using almost the same set of data. Note that all these groups
have used the GRV unpolarized parton densities \cite{GRV} for
constraining their polarized parton densities at large $x$. In
this $x$ region the unpolarized GRV and MRST'02 gluons are
practically the same, while the magnitude of the unpolarized GRV
strange sea quarks is much smaller than that of MRST'02.
Therefore, the GRV unpolarized strange sea quarks provide a
stronger constraint on the polarized ones (see Fig. 3). The
impact on the determination of the polarized strange sea density
is demonstrated in Fig. 3. As a result, the magnitude of our
polarized strange sea density $x\vert {\Delta s(x, Q^2)}\vert$ is
larger in the region $x> 0.1$ than those obtained by the other
groups. Note also that the magnitude of $x\Delta s$ obtained by
the GRSV and BB is smaller than that determined by AAC. We
consider the GRSV result to be a consequence of the fact that in
their analysis, the GRV positivity constraint is imposed at lower
value of $Q^2$: $Q^2= {\mu}^2_{NLO}=0.4~GeV^2$, while AAC has
used the same requirement at $Q^2=1~GeV^2$. Finally, the
different positivity conditions on $\Delta s$ influence also the
determination of the polarized gluon density for larger $Q^2$
because the evolution in $Q^2$ mixes the polarized sea quarks and
gluons.
\begin{figure}
\resizebox{.4\columnwidth}{!}
  {\includegraphics{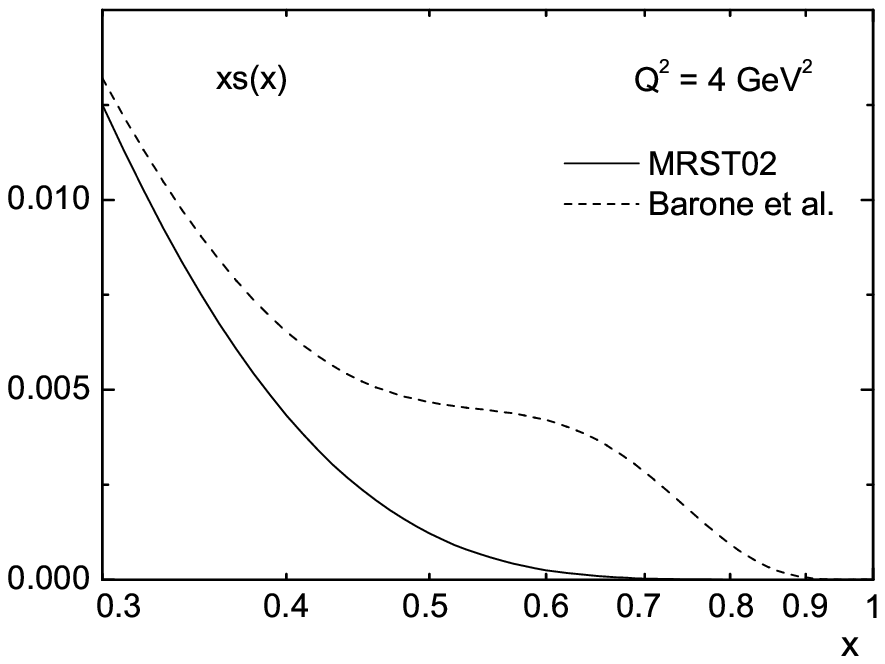}}
\resizebox{.4\columnwidth}{!}
    {\includegraphics{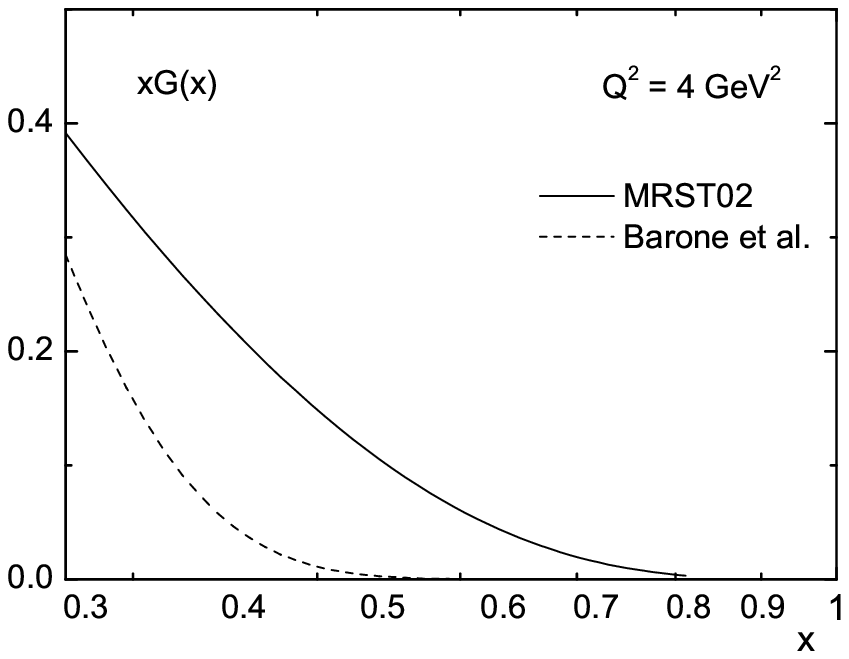}}
  \caption{Comparison between the NLO(${\rm \overline {MS}}$) unpolarized
strange quark sea and gluon densities determined by MRST'02
\cite{MRST02} and Barone at al. \cite{Barone}.}
\end{figure}

To end this discussion, we would like to emphasize that for the
adequate determination of polarized strange quarks and gluons at
large $x$, the role of the corresponding unpolarized densities is
very important. That is why the latter have to be determined with
good accuracy at large $x$ in the preasymptotic $(Q^2,~ W^2)$
region too. Usually the sets of unpolarized parton densities,
presented in the literature, are extracted from the data on DIS
using cuts in $Q^2$ and $W^2$ chosen in order to minimize the
higher twist effects. In order to use the densities for
constraining the polarized parton densities they have to be
continued to the preasymptotic $(Q^2,~W^2)$ region. It is not
obvious that the continued unpolarized parton densities would
coincide well with those obtained from the data in the region
$(Q^2>1~GeV^2,~ W^2 > 4~GeV^2)$ in the presence of the HT
corrections to unpolarized structure functions $F_1$ and $F_2$.
So, a QCD analysis of the unpolarized world data including the
preasymptotic $(Q^2,~W^2)$ region and taking into account HT
corrections is needed in order to extract correctly the
unpolarized parton densities in the preasymptotic region. Our
arguments for the need for a precise determination of the
unpolarized densities of strange quarks and gluons in both the
asymptotic and preasymptotic regions in $Q^2$ and $W^2$, coming
from spin physics, could be considered as additional to those
discussed in the recent paper \cite{Forte}.
\begin{figure}
  \includegraphics[height=.26
  \textheight]{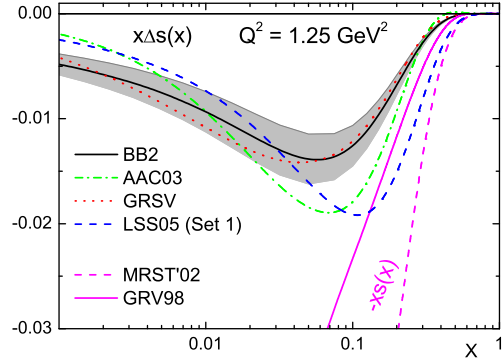}
  \caption{Comparison between our NLO(${\rm \overline {MS}}$)
polarized strange sea quark density (Set 1) \protect\cite{LSS05}
at $Q^2=1.25~GeV^2$ with those \protect\cite{Groups} obtained by
GRSV ('standard scenario'), BB (ISET=4 or BB2) and AAC (AAC03).
The unpolarized MRST02 and GRV98 strange sea quark densities are
also shown.}
\end{figure}
\begin{theacknowledgments}
This research was supported by the JINR-Bulgaria Collaborative
Grant, by the RFBR (No 05-01-00992, 03-02-16816), by the
Bulgarian National Science Foundation under Contract Ph-1010 and
by the Royal Society of Edinburgh Auber Bequest.
\end{theacknowledgments}

\end{document}